\documentclass{emulateapj}  
  \revised{}
  \accepted{}

\slugcomment{Submitted to \apj}
\shortauthors{Auger et al.}
\shorttitle{Lens Galaxy Properties of SBS1520+530}

\newcommand{\kms}{km\ s$^{-1}$}
\newcommand{\mkms}{\rm{km s^{-1}}}
\newcommand{\ang}{$\rm{\AA}$}
\newcommand{\hinv}{h$^{-1}$}
\newcommand{\hunit}{~km\ s$^{-1}$\ Mpc$^{-1}$}

\received{2007 June 27}
\begin{document}

\title{Lens Galaxy Properties of SBS1520+530: Insights from Keck Spectroscopy and AO Imaging}

\author{M. W. Auger, C. D. Fassnacht, K. C. Wong}
\affil{
   Department of Physics, University of California, 1 Shields Avenue,
   Davis, CA 95616, USA }
\email{mauger@physics.ucdavis.edu, fassnacht@physics.ucdavis.edu}
\author{D. Thompson}
\affil{
   Large Binocular Telescope Observatory, University of Arizona, 933 N. Cherry Avenue, Tucson, AZ  85721, USA
}

\author{K. Matthews, B. T. Soifer}
\affil{
   Caltech Optical Observatories, California Institute of Technology, Pasadena, CA 91125, USA
}

\begin{abstract}
We report on an investigation of the SBS 1520+530 gravitational lens system. We have used archival HST imaging, Keck spectroscopic data, and Keck adaptive-optics imaging to study the lensing galaxy and its environment. The AO imaging has allowed us to fix the lens galaxy properties with a high degree of accuracy when performing the lens modeling, and the data indicate that the lens has an elliptical morphology and perhaps a disk. The new spectroscopic data suggest that previous determinations of the lens redshift may be incorrect, and we report an updated, though inconclusive, value $z_{lens} = 0.761$. We have also spectroscopically confirmed the existence of several galaxy groups at approximately the redshift of the lens system. We create new models of the lens system that explicitly account for the environment of the lens, and we also include improved constraints on the lensing galaxy from our adaptive-optics imaging. Lens models created with these new data can be well-fit with a steeper than isothermal mass slope ($\alpha = 2.29$, where $\rho \propto r^{-\alpha}$) if $H_0$ is fixed at 72\hunit; isothermal models require $H_0 \sim 50$\hunit.  The steepened profile may indicate that the lens is in a transient perturbed state caused by interactions with a nearby galaxy.
\end{abstract}


\section{INTRODUCTION}
The strong gravitational lens system SBS 1520+530 (hereafter SBS1520) was first investigated by \citet{chavushyan}, who found that the system consists of a pair of images of a broad absorption line quasar ($z_{src} = 1.855$) separated by 1\farcs6. The lensing galaxy was soon detected with adaptive optics (AO) imaging \citep{crampton} and was assumed to be at the redshift of one of two absorption line systems seen in the spectra of the quasar images. \citet{burud} attempted to deconvolve the lens spectrum from the quasar spectra and found the redshift of the lens to be $z_{lens} = 0.717$. This redshift is broadly consistent with a photometric redshift determined by \citet{faure}. Furthermore, the lens was found to lie along the line of sight to a photometrically identified cluster of galaxies that is expected to be at approximately the same redshift as the lens \citep{burud,faure}.

Optical monitoring campaigns have led to the measurement of a time delay between the quasar images of $\sim 130$ days \citep{burud,khamitov}. This time delay provides an additional constraint for determining the mass slope of the lens galaxy \citep[e.g.,][]{rusin} and allows a value of the Hubble Constant to be determined for the system \citep[$H_0 = 51$\hunit assuming an isothermal mass profile;][]{burud}. Note, however, that the mass slope and, thus, the value of $H_0$ depend on the environment surrounding the lens system \citep[e.g.,][]{dobke,auger}. An incorrect understanding of the mass distribution and environment of the lens might account for the anomalously low value of $H_0$ obtained for SBS1520 compared to other lens systems \citep[e.g.,][]{koopmans03,york} and the WMAP \citep[][]{spergel} and \em{Hubble Space Telescope} (\em{HST}) Key Project \citep[][]{freedman} results.

In this paper we present new Keck AO and archival \emph{HST} imaging of the lens system that indicates the lensing galaxy may have a disk component. We also present a spectroscopic investigation of the lens environment and provide a new analysis of the lensing galaxy's spectrum which results in an updated lens redshift of $z_{lens} = 0.761$. We discuss the implications of these new observational data on previous analyses performed with SBS1520 and suggest that, in spite of some complexity, this lens provides an interesting platform to investigate dark matter interactions in dense environments.

\section{IMAGING}
High resolution imaging of SBS1520 at 2.12~$\mu$m ({\it K$_p$}-band) was obtained on 2006 May 2 with the Near InfraRed Camera 2 (NIRC2; Matthews et al. in preparation) behind the adaptive-optics bench on the W. M. Keck-II Telescope. The NIRC2 narrow camera was used for the imaging, which provides a field of view of 10\arcsec~$\times$~10\arcsec~and a pixel-scale of 9.94~mas~pixel$^{-1}$. The data were taken in nine 90-s exposures with a small dither between each exposure to facilitate good sky background subtraction during the reduction process. A sodium laser guide star was used to correct for the atmospheric turbulence, and a $r = 12$ star, 15\arcsec~from SBS1520, was used for Tip Tilt correction.

The data were reduced within {\sc iraf}\footnote{{\sc iraf} (Image Reduction and Analysis Facility) is distributed by the National Optical Astronomy Observatories, which are operated by AURA, Inc., under cooperative agreement with the National Science Foundation.} using a double-pass reduction algorithm. The first pass is used to determine the positions of the objects on the sky, which are then masked out during the second pass where the usual sky subtraction with temporally adjacent frames is carried out. Bad pixel and cosmic ray masks are generated during this process and these masks are then used to construct weight maps. The individual exposures are combined using the weight maps and the {\it drizzle} package \citep{fruchter02}, which also corrects the data for the geometric distortions across the NIRC2 camera to provide precise relative astrometry. The point-source full-width at half maximum in the final drizzled image is 54 milliarcseconds.

\begin{figure*}[ht]
\epsscale{1}
\plotone{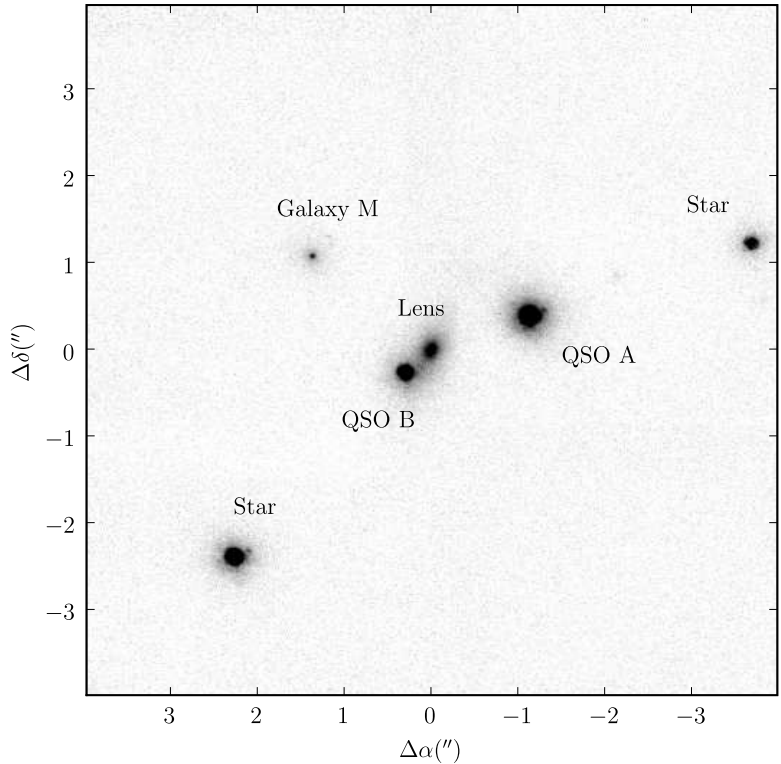}
  \caption{Keck NIRC2 adaptive optics K-band image of the SBS1520 lens system. The image is centered on the lensing galaxy, and two field stars are indicated. The galaxy to the north east of the lens is denoted Galaxy M. The lensed quasar components are also indicated. The resolution of the image is 0\farcs05.}
\label{figure_1520_lens}
\end{figure*}

\begin{figure*}[ht]
\begin{center}
\epsscale{1}
 \includegraphics[width=0.85\textwidth]{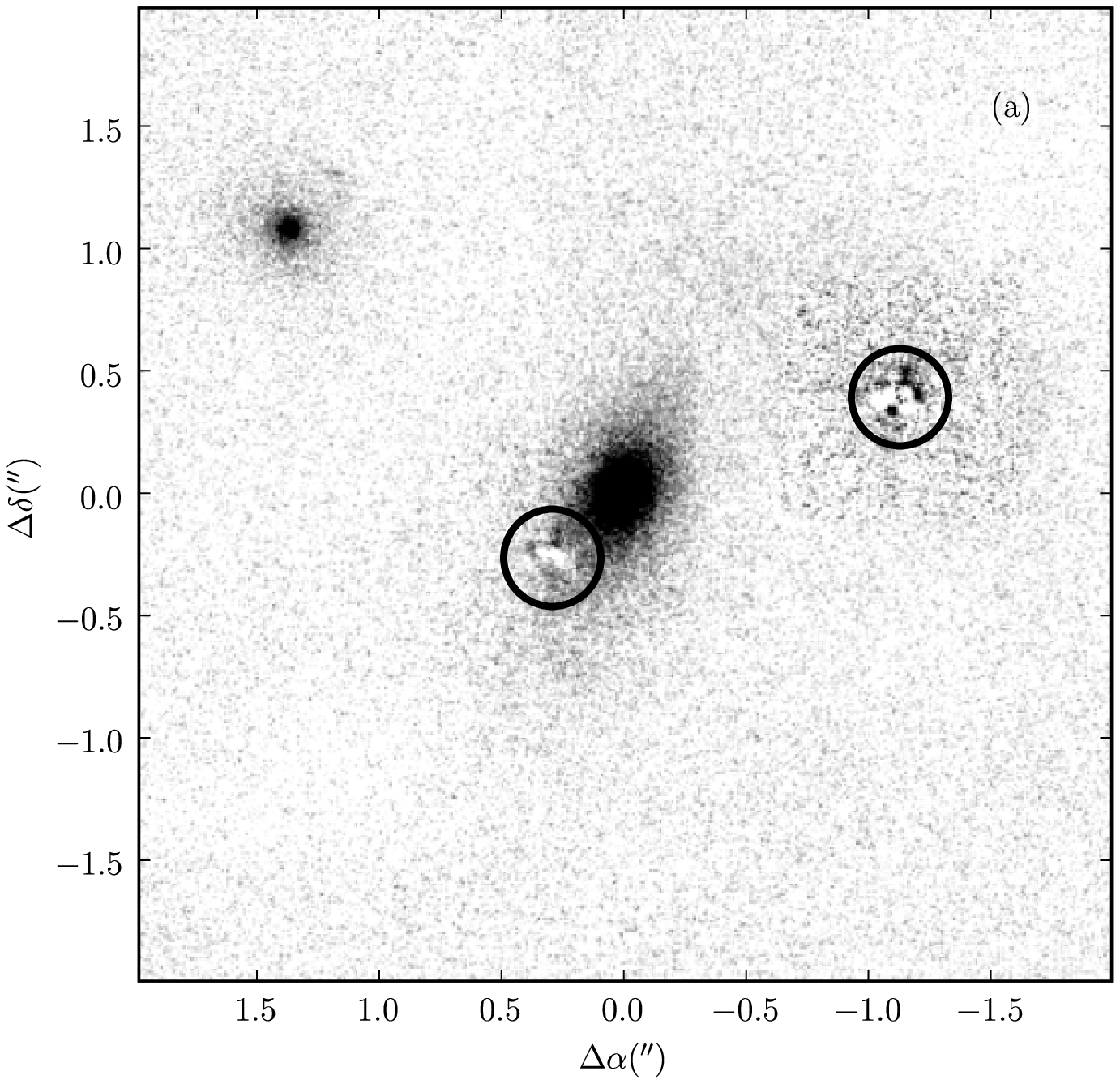}
 \includegraphics[width=0.85\textwidth]{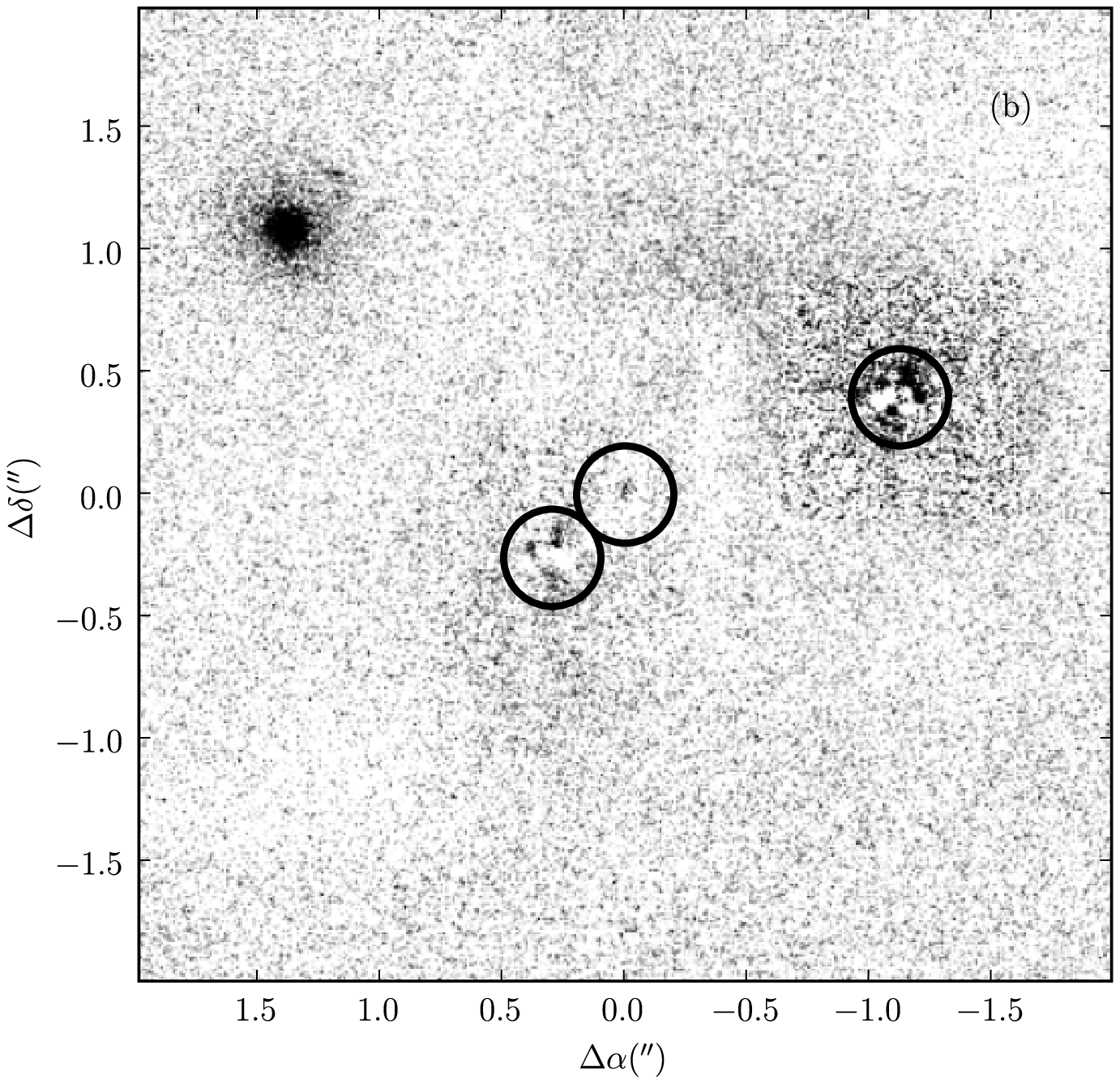}
  \caption{Keck AO K-band image of SBS1520 with (a) the quasar images subtracted and (b) the quasar images and the Sersic-model of the lensing galaxy subtracted. Circles indicate the centers of the subtracted components; note that a faint core is noticeable in the residuals of the galaxy.}
\label{figure_subtracted}
\end{center}
\end{figure*}

The inner 8\arcsec~of the AO imaging is shown in Figure \ref{figure_1520_lens}, clearly showing the two lensed quasar images, the lensing galaxy, and a nearby galaxy. The high resolution and sensitivity of the observations allow us to determine the morphological parameters of the lensing galaxy with a high degree of confidence. The A and B images of the quasar were fitted and subtracted using an empirical PSF determined from the star to the southeast of the lens (Figure \ref{figure_subtracted}a); the K-band image of the lens shows an early type morphology, perhaps with a disk. A best-fit galaxy model was then determined for the lensing galaxy using a non-linear fitting code to model the lens with a Sersic model convolved with the empirical PSF. The residuals from this model are shown in Figure \ref{figure_subtracted}b; the best-fit Sersic index is $n = 2.7$ and the effective radius is determined to be $r_e = 0\farcs49$. The ellipticity from the model is found to be 0.44 at a position angle of 156\arcdeg.

\begin{figure*}[ht]
\epsscale{1}
\plotone{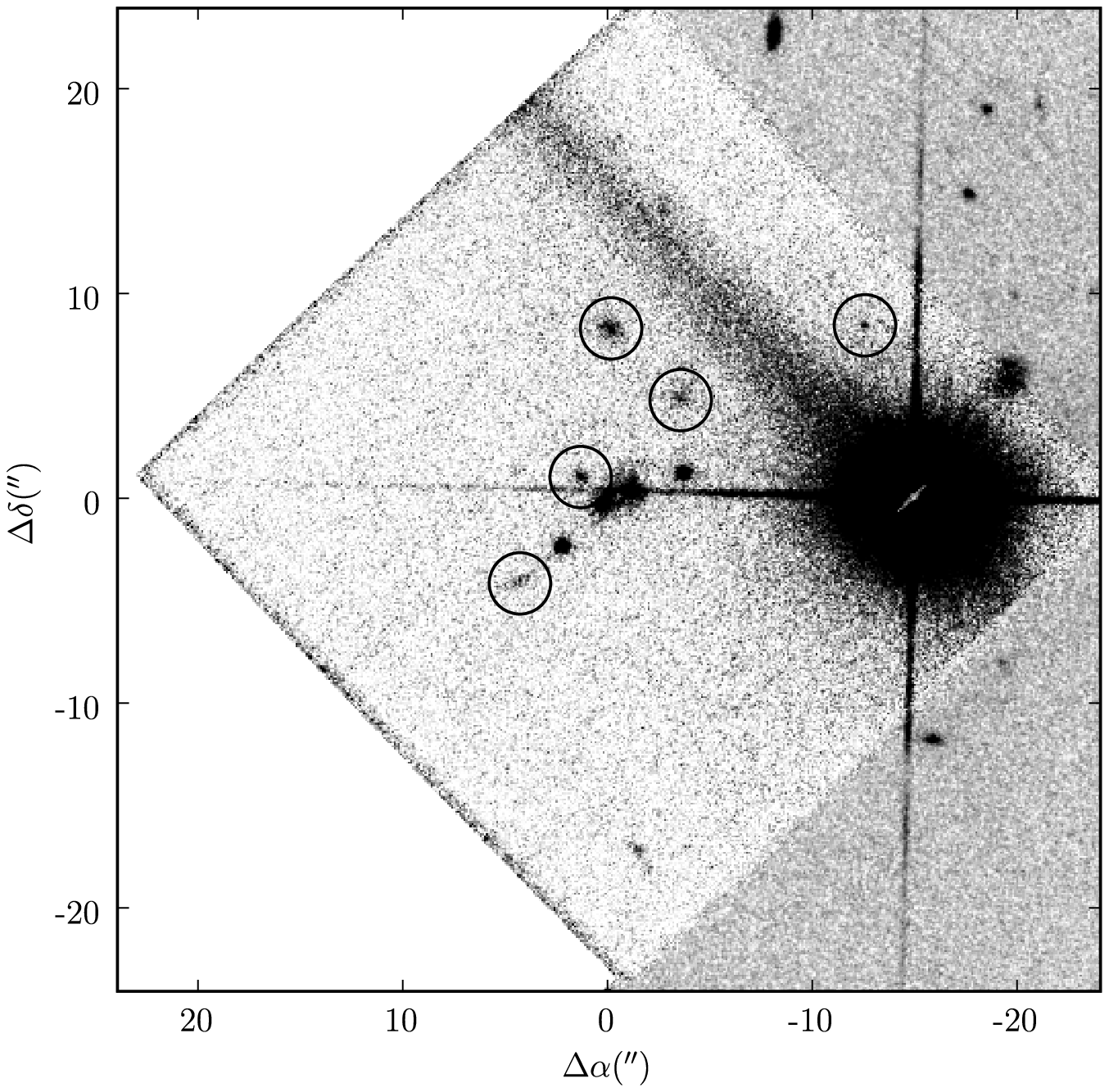}
  \caption{WFPC2 F814W image of the SBS1520 field. All galaxies within 16\arcsec~of the lens are indicated with circles and have been explicitly included in the lens modeling. The bright star 15\arcsec~west of the lens was used for the adaptive optics imaging.}
\label{figure_f814w}
\end{figure*}

We have also obtained \emph{HST} WFPC2 and NICMOS archival data for SBS1520 (\emph{HST} GO Programs 8175, 7495, and 7887; PI Falco). The imaging data consist of 2100 s with the F555W filter and 1600 s with F814W using WFPC2, and 2816 s with F160W and the NIC2 camera. These data were reduced using the {\em multidrizzle}\footnote{{\em multidrizzle} is a product of the Space Telescope Science Institute, which is operated by AURA for NASA.} package \citep{koekemoer}, and galaxy properties were measured using SExtractor \citep{bertin}. The relative positions of the sources used in our analysis have been measured from the NIRC2 imaging if the objects are in the NIRC2 field of view; otherwise, the F814W imaging has been used. The lens system was placed at the center of the PC chip for the WFPC2 data and in the center of the NICMOS field, providing complete coverage out to 16\arcsec~for the WFPC2 fields (see Figure \ref{figure_f814w}) and 12\arcsec~for the NICMOS field. The object $\sim 1\farcs8$ to the Northeast of the lens is not resolved in conventional ground-based imaging but is revealed to be a galaxy in AO and \emph{HST} imaging, with a FWHM approximately 30\% larger than the PSF. Following \citet{burud}, we will denote this galaxy as Galaxy M.

\section{SPECTROSCOPY}
We have obtained spectroscopic observations of SBS1520 using the Echellette Spectrograph and Imager \citep[ESI;][]{sheinis} on the Keck II telescope and the Low-Resolution Imaging Spectrograph \citep[LRIS;][]{oke} on the Keck I telescope. Our primary lens spectroscopy was obtained with ESI on 2000 July 3 in conditions with thin cirrus and approximately 0\farcs8 seeing. Four 1800 s exposures were observed with a 0\farcs75 slit. The ESI slit was oriented to run through both quasar images and through the lensing galaxy. Two LRIS slitmask observations were made on 2004 June 15 in non-photometric conditions with approximately 1\farcs5 seeing. The purpose of the LRIS observations was to measure redshifts for both the lensing galaxy and nearby galaxies. The 831/9200 grating, with dispersion 0.9~\ang~pixel$^{-1}$, was used on the red arm of the spectrograph and the blue arm employed the 600/4000 grism with dispersion 0.63~\ang~pixel$^{-1}$. Four exposures of 1800 s were obtained for each mask, though one exposure of the second mask was discarded due to poor transparency. Both of the slitmasks contained a slit placed over the lens system in the same orientation as the ESI slit, providing seven good LRIS exposures of 1800 s for the lens system.

\subsection{Data Reduction}
The LRIS data were reduced using a custom automated pipeline written in Python (Auger, in preparation). The pipeline removes the instrumental bias and overscan regions of the LRIS CCDs, corrects for amplifier gain offsets, and flat fields the data from the red arm of the spectrograph. The program then automatically determines a wavelength solution from arclamp exposures and skylines, corrects for telluric absorption features with an airmass-scaled model of the typical absorption at Keck, performs an `optimal' background subtraction \citep{kelson}, and resamples and coadds the two-dimensional spectra to a constant wavelength-scale grid. The pipeline also searches each slit for spectral traces or emission lines and performs an aperture-weighted extraction. The extracted spectra were cross-correlated with galactic and stellar template spectra to determine redshifts for the science targets. All of these redshifts were then verified by hand; several were discarded due to poor quality spectra while others were corrected by manually identifying features.

The ESI data were reduced with a modified version of the Python scripts used to reduce the LRIS data. A star was observed at several positions along the slit and the traces of the star were used to straighten the echelle dispersion orders. A wavelength solution was determined from HgNe, CuAr, and Xe arclamp observations. The science data and an observation of the spectrophotometric standard star HZ44 \citep{hz44} were then straightened and the night sky background was subtracted. The spectrum of HZ44 was then extracted and an atmospheric absorption and instrumental response model was created. The SBS1520 quasar spectra were then traced and extracted. The galaxy spectrum was extracted by defining a Gaussian aperture with a full-width at half maximum of one pixel centered slightly off of the position of the lens galaxy to minimize the shot noise from the quasar light. The extracted spectra from the four exposures were then coadded and a smoothed version of the galaxy spectrum was created by convolving the galaxy spectrum with a Gaussian kernel of width $\sigma = 100~\mkms$. We also extracted the galaxy spectrum by modeling the quasar images and lens galaxy in a method similar to \citet{burud} \citep[also see][]{courbin,eigenbrod}. Two stars (shown in Figure \ref{figure_1520_lens}) were observed on the ESI and LRIS slits along with the lens system to help define the PSF, but we found that our observations did not have sufficient sensitivity to adequately model the galaxy and quasar spectra without substantial artifacts.

\subsection{SBS1520 Quasar and Lens Spectra}
\label{section_lens_redshift}
\begin{figure*}[ht]
\epsscale{1}
\plotone{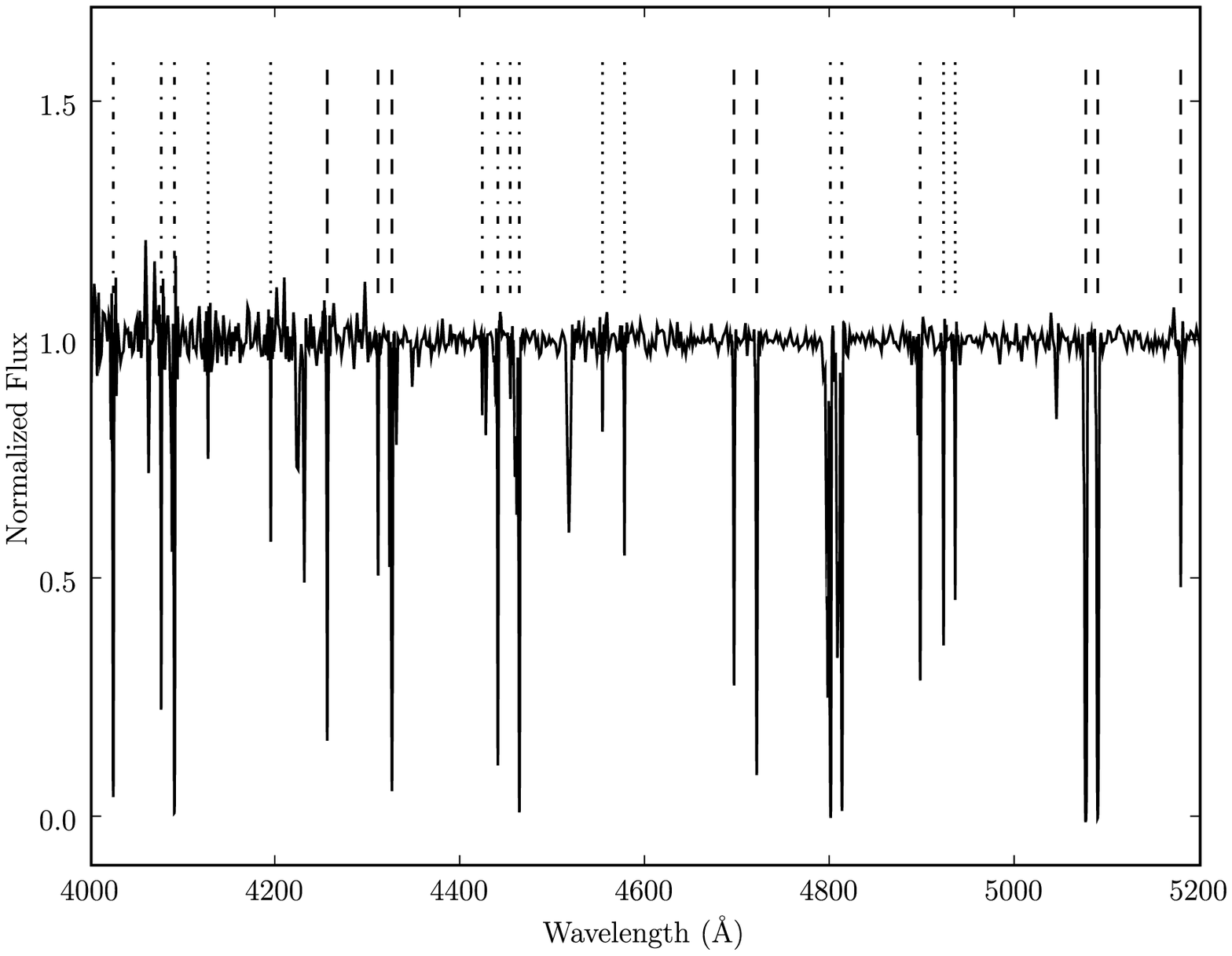}
  \caption{There are four absorption systems seen in the SBS1520 quasar spectra; a portion of the extracted spectrum from the A component is shown above. Prominent features from the $z = 0.717$ (dash-dotted), $z = 0.760$ (dotted), and $z = 0.815$ (dashed) systems are indicated; most of these features are Fe and Mg lines. The spectrum has been divided by the continuum and several of the MgII lines are seen to be saturated.}
\label{figure_absorption}
\end{figure*}

Previous observations of SBS1520 have revealed several line-of-sight absorption systems present in the quasar spectra at $z \approx 0.717$ and $z \approx 0.815$ \citep{chavushyan,burud}. \citet{burud} found evidence for CaII K \& H absorption features associated with the $z = 0.717$ system in the two quasar spectra and also find these lines in a spectrum of the lens galaxy obtained by deconvolving the lens spectrum from the quasar spectra. The absence of CaII features associated with the $z = 0.815$ system led those authors to assign the lens galaxy the redshift $z = 0.717$. Our data confirm the absorption systems seen at $z = 0.717$ and $z = 0.815$, but the quasar spectra also reveal a Mg and Fe absorption system at $z = 0.760$ and a CIV absorption system at $z = 1.728$ (Figure \ref{figure_absorption}). Furthermore, we have also found evidence for a galaxy at $z = 0.761$ that might be the lensing galaxy, with the $z = 0.717$ galaxy being a local perturber. The three lower redshift absorption systems are potentially associated with the lensing galaxy, and we evaluate the likelihood that the lens is at each of these redshifts.

\subsubsection{$z = 0.815$ Candidate}
Previous studies did not find evidence for a stellar population associated with the $z = 0.815$ absorption system \citep{burud} and we are also unable to find any broad absorption features that might be associated with the stellar population of a massive lensing galaxy; there were no features associated with the $z = 0.815$ system in our extracted galaxy spectrum. The quasar spectra reveal a Mg and Fe absorption system that also contains narrow CaII H \& K absorption lines, though these CaII features are not associated with a stellar population. Furthermore, we note that the absorption lines are stronger in the spectrum of quasar image A than quasar image B, which is contrary to expectation because the galaxy is much closer to image B (see Figure \ref{figure_1520_lens}). A tabulation of the absorption features at $z = 0.815$ is found in Table \ref{table_0815_lines}. The absence of a stellar component and the stronger features being present in the spectrum of image A make it highly unlikely that the primary lensing galaxy is associated with this absorption system. Strong MgII absorption systems have been observed at high impact parameter \citep{churchill} and there are several galaxies within 16\arcsec~(approximately 80~\hinv~kpc at $z = 0.815$) of the lens that could be responsible for the quasar absorption lines.

\begin{deluxetable*}{lccccc}
\tabletypesize{\scriptsize}
\tablecolumns{6}
\tablewidth{0pc}
\tablecaption{QSO Absorption at $z = 0.8154$}
\tablehead{
 &
 \multicolumn{2}{c}{Component A} &&
 \multicolumn{2}{c}{Component B} \\
 \cline{2-3} \cline{5-6} 
 \colhead{Line} &
 \colhead{REW (\ang)} &
 \colhead{Wavelength (\ang)} &&
 \colhead{REW (\ang)} &
 \colhead{Wavelength (\ang)}
}
\startdata
  FeII2344 & 0.58 & 4255.448 && 0.34 & 4255.589 \\
  FeII2374 & 0.29 & 4310.385 && 0.15 & 4310.473 \\
  FeII2382 & 1.20 & 4325.458 && 0.75 & 4325.630 \\
  FeII2586 & 0.52 & 4695.580 && 0.30 & 4695.703 \\
  FeII2600 & 0.87 & 4720.156 && 0.69 & 4720.290 \\
  MgII2796 & 1.68 & 5076.165 && 1.35 & 5076.384 \\
  MgII2803 & 1.42 & 5089.242 && 1.09 & 5089.421 \\
   MgI2852 & 0.41 & 5179.019 && 0.25 & 5179.208 \\
  CaII3934 & 0.12 & 7142.672 && \nodata & \nodata \\
  CaII3969 & 0.14 & 7205.966 && \nodata & \nodata \\
\enddata
\label{table_0815_lines}
\end{deluxetable*}

\subsubsection{$z = 0.717$ Candidate}
\begin{figure*}[ht]
\epsscale{1}
\plotone{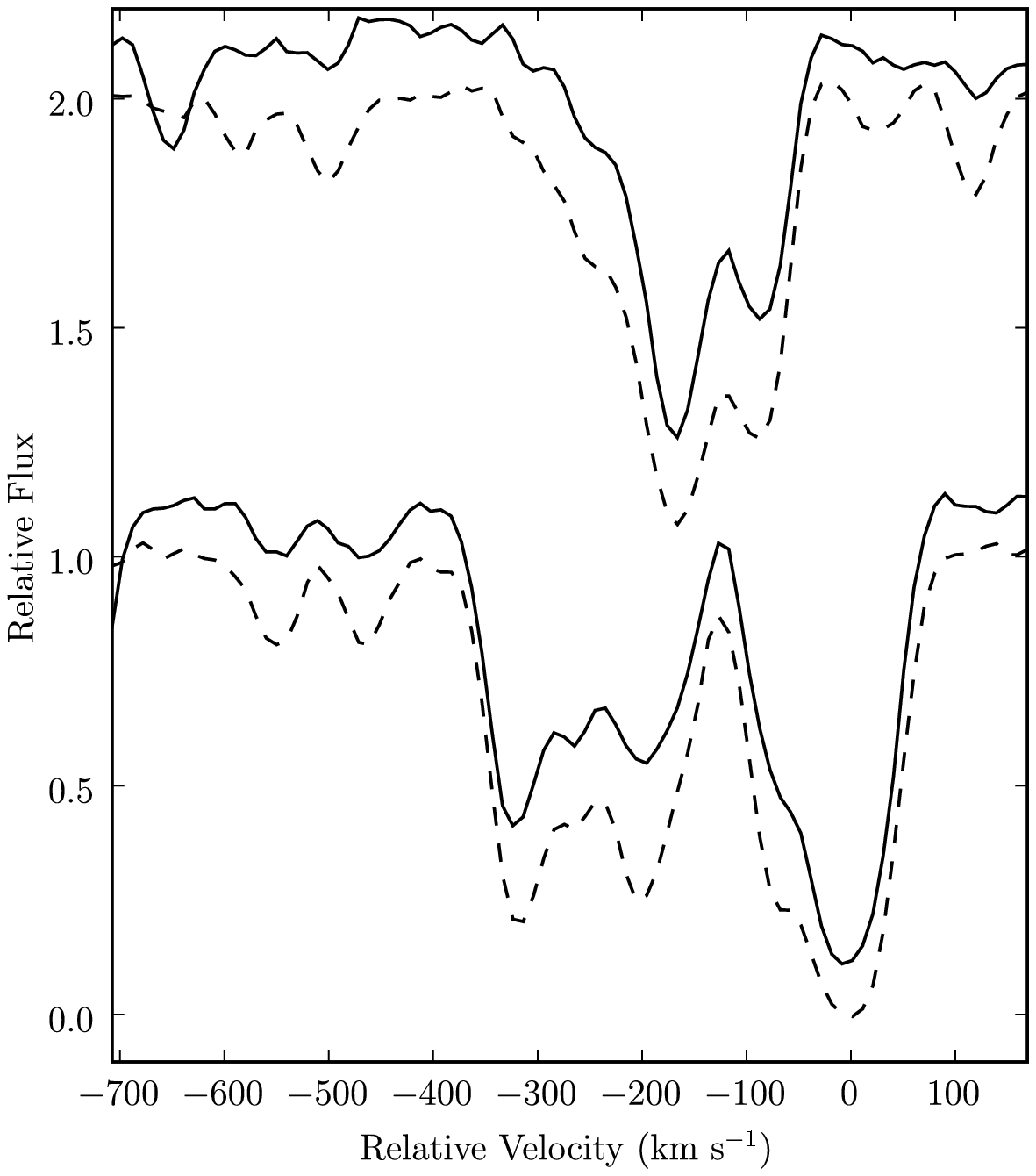}
  \caption{The MgII doublet at 2796\ang~(dashed lines) and 2803\ang~(solid lines, offset by 0.1 for clarity) for the $z = 0.717$ absorption system. The lines from the B component have been shifted upward by one unit for clarity and the zero point for the velocity has been taken to be the center of the deepest feature in the spectrum of the A image.}
\label{figure_mgii_0717}
\end{figure*}

The SBS1520 quasar spectra also show very clear evidence of CaII absorption at a redshift $z = 0.717$ (Table \ref{table_0717_lines}) associated with a system of Mg and Fe absorption. The structure of this absorption system's MgII lines at 2796\ang~and 2803\ang~is complex and contains features that span a velocity range of approximately 550~\kms; this large spread in the velocities of the substructure components might be indicative of infall in a forming galaxy. The spectrum from the B image of the quasar does not show as much substructure as the A component, though the velocity spread is essentially the same as the A component (Figure \ref{figure_mgii_0717}). The absorption lines are much more prominent in the spectrum of quasar image A compared to quasar image B; this is counter to what might be expected if the absorption system is associated with the lens galaxy because the lensing galaxy is closer to image B.

\begin{figure*}[ht]
\epsscale{1}
\plotone{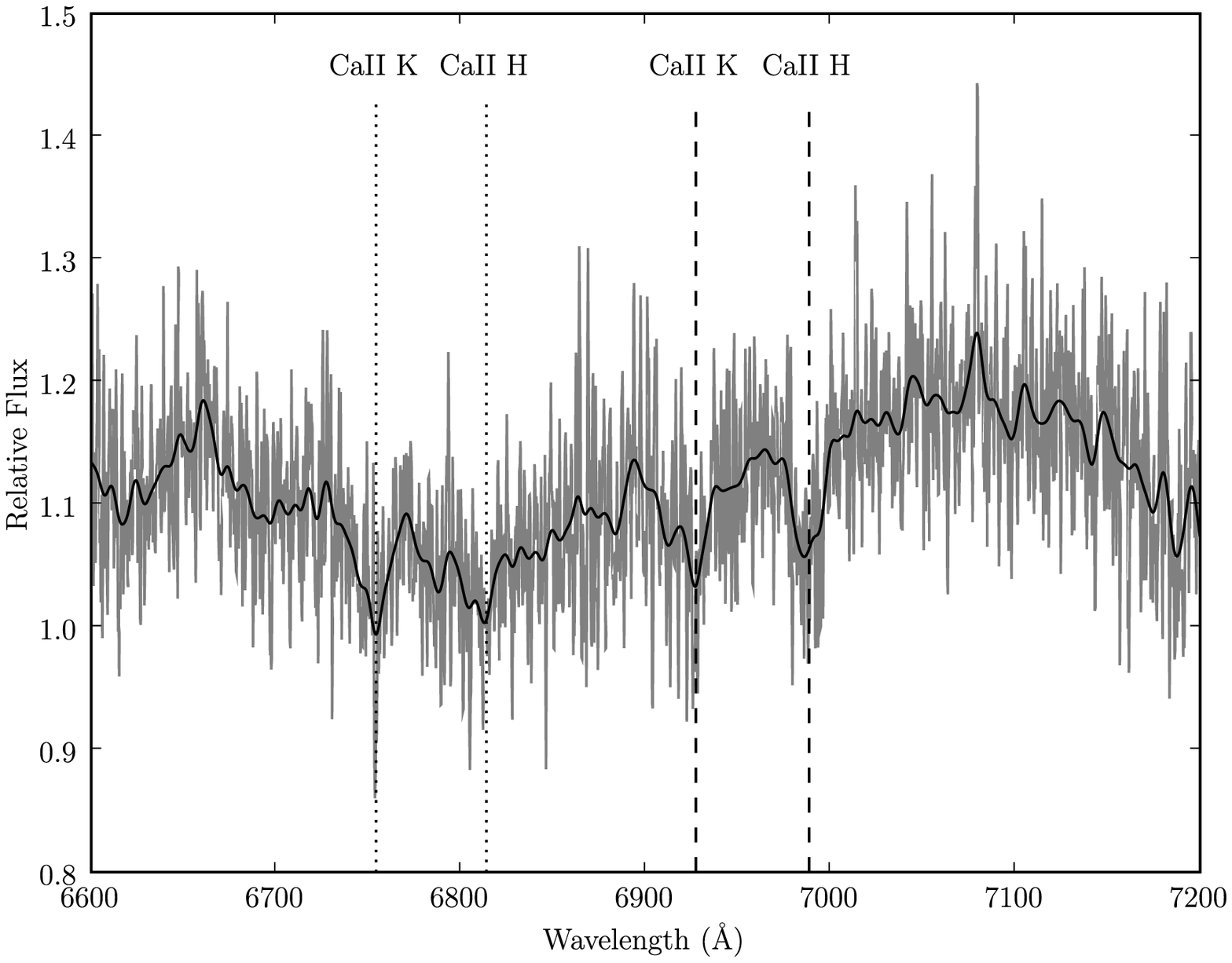}
  \caption{ESI spectrum of the lensing galaxy of SBS1520 between 6600\ang~and 7200\ang. CaII K \& H from the lens galaxy at $z = 0.761$ are marked with dashed lines; features from an intervening absorption system at $z = 0.717$ are marked with dotted lines. The spectrum, smoothed with a Gaussian kernel of width $\sigma = 100~\mkms$, is overplotted.}
\label{figure_lens_zspec}
\end{figure*}

The spectrum of the lens galaxy, extracted from an aperture between the quasar spectra, displays the broad CaII absorption features at $z = 0.717$ identified by \citet{burud}. The lines are weak (see Figure \ref{figure_lens_zspec}) but are at the same redshift as the absorption system. The presence of these lines led \citet{burud} to ascribe the redshift of this system to the lensing galaxy. However, evidence against this identification includes the stronger absorption features in the A component compared to the B component, the large amount of substructure seen in the A component indicative of a forming galaxy, and the presence of a stronger set of stellar absorption features at a redshift $z = 0.761$.

\begin{deluxetable*}{lccccc}
\tabletypesize{\scriptsize}
\tablecolumns{6}
\tablewidth{0pc}
\tablecaption{QSO Absorption at $z = 0.7165$}
\tablehead{
 &
 \multicolumn{2}{c}{Component A} &&
 \multicolumn{2}{c}{Component B} \\
 \cline{2-3} \cline{5-6}
 \colhead{Line} &
 \colhead{REW (\ang)} &
 \colhead{Wavelength (\ang)} &&
 \colhead{REW (\ang)} &
 \colhead{Wavelength (\ang)}
}
\startdata
  FeII2344 & 0.80 & 4023.84 && 0.25 & 4021.76 \\
  FeII2374 & 0.42 & 4075.75 && 0.07 & 4073.63 \\
  FeII2382 & 1.24 & 4090.04 && 0.59 & 4088.17 \\
  MnII2576 & 0.18 & 4423.17 && \nodata & \nodata \\
  FeII2586 & 0.70 & 4440.07 && 0.30 & 4437.73 \\
  MnII2594 & 0.23 & 4453.47 && \nodata & \nodata \\
  FeII2600 & 1.38 & 4463.29 && 0.65 & 4461.24 \\
  MgII2796 & 2.58 & 4799.89 && 1.49 & 4797.73 \\
  MgII2803 & 2.13 & 4812.28 && 0.94 & 4810.11 \\
   MgI2852 & 0.69 & 4897.08 && 0.13 & 4894.67 \\
  CaII3934 & 0.45 & 6753.93 && 0.08 & 6750.38 \\
  CaII3969 & 0.25 & 6813.67 && \nodata & \nodata \\
\enddata
\label{table_0717_lines}
\end{deluxetable*}

\subsubsection{$z = 0.761$ Candidate}
Another pair of broad CaII lines are found in the galaxy spectrum at 6986.64\ang~and 6926.06\ang~($z = 0.7606$, Figure \ref{figure_lens_zspec}). The CaII H feature from this system has a line strength nearly two times greater than the CaII H line from the $z = 0.717$ system. The line at 6986\ang~might also be seen in the spectrum of quasar B in Figure 7 of \citet{burud}, though we do not find evidence for the CaII lines in our spectra of either of the quasar images. There is an Fe and Mg absorption system discovered in our ESI and LRIS quasar spectra (Table \ref{table_0760_lines}) at the redshift $z_A = 0.7603$. The absorption features in the spectrum of quasar image B have greater rest equivalent widths than the corresponding features in the spectrum of image A (Table \ref{table_0760_lines}), suggesting that the absorption system is closer to image B than image A. There is also a velocity offset of $v_{B-A} \approx 200\mkms$ between the absorption seen in the A and B images of the quasar; this suggests that there may be two galaxies associated with the cluster at $z = 0.76$ (see Section \ref{section_lens_environment}) responsible for the quasar absorption spectra or there may be a rotational component associated with the possible disk seen in the AO imaging (Figure \ref{figure_subtracted}a).

\begin{deluxetable*}{lccccc}
\tabletypesize{\scriptsize}
\tablecolumns{6}
\tablewidth{0pc}
\tablecaption{QSO Absorption at $z = 0.7603$}
\tablehead{
 &
 \multicolumn{2}{c}{Component A} &&
 \multicolumn{2}{c}{Component B} \\
 \cline{2-3} \cline{5-6}
 \colhead{Line} &
 \colhead{REW (\ang)} &
 \colhead{Wavelength (\ang)} &&
 \colhead{REW (\ang)} &
 \colhead{Wavelength (\ang)}
}
\startdata
  FeII2344 & 0.07 & 4126.442 && 0.08 & 4129.185 \\
  FeII2382 & 0.18 & 4194.293 && 0.25 & 4196.938 \\
  FeII2586 & 0.08 & 4553.237 && 0.07 & 4556.341 \\
  FeII2600 & 0.21 & 4577.002 && 0.14 & 4580.098 \\
  MgII2796 & 0.38 & 4922.277 && 0.55 & 4924.809 \\
  MgII2803 & 0.31 & 4934.881 && 0.42 & 4938.293 \\
\enddata
\label{table_0760_lines}
\end{deluxetable*}

\subsection{Lens Environment Spectra}
\label{section_lens_environment}
The two LRIS slitmask observations of the field of SBS1520 have yielded redshift identifications for 52 galaxies. We treat the lens galaxy as being at a redshift $z = 0.761$ and use the objective group finding algorithm described in \citet{auger} to find groups in the field of SBS1520. To summarize that algorithm, we define a maximum rest-frame velocity difference between galaxies in a candidate group; galaxies with velocity differences larger than this are placed into a new candidate group. We then define a mean group center and redshift for each group candidate, determine a velocity dispersion for the group using the gapper method, and apply an iterative clipping algorithm to remove outliers. This procedure leads to the identification of three spectroscopically confirmed groups in the SBS1520 field, each associated with one of the absorption systems described in Section \ref{section_lens_redshift}. In Table \ref{table_lens_groups} we describe the group properties, including the mean position, the number of galaxies in the group, the offset of the group centroid from the lensing galaxy, the redshift, velocity dispersion, and virial radius of the group, and the convergence ($\kappa$) produced by the group at the location of the lens assuming a singular isothermal sphere (SIS) mass profile for the group. Table \ref{table_group_members} contains information for the galaxies that populate each group.

\begin{deluxetable*}{llcccccc}
\tabletypesize{\scriptsize}
\tablecolumns{8} 
\tablewidth{0pc} 
\tablecaption{SBS1520 Groups}
\tablehead{
 \multicolumn{2}{c}{Mean Position} & 
 &
 \colhead{Lens Offset} &
 &
 \colhead{$\sigma$} & 
 \colhead{R$_{vir}$}
 & \\
 \colhead{RA} &
 \colhead{Dec} &
 \colhead{N$_{gals}$} &
 \colhead{(\arcsec)} &
 \colhead{z} &
 \colhead{\kms} & 
 \colhead{(\hinv~Mpc)} &
 \colhead{$\kappa$}
}
\startdata
15 21 48.16 & 52 53 19.5 &  5 &  94 & 0.716 & $121 \pm 32$ & 0.92 & < 0.005 \\
15 21 38.22 & 52 55 23.1 & 13 &  69 & 0.758 & $465 \pm 86$ & 0.45 & 0.02 \\
15 21 43.37 & 52 54 52.0 & 10 &  14 & 0.818 & $369 \pm 118$ & 0.87 & 0.06 \\
\enddata
\label{table_lens_groups}
\end{deluxetable*}

\begin{deluxetable*}{llllccc}
\tabletypesize{\scriptsize}
\tablecolumns{7}
\tablewidth{0pc}
\tablecaption{Group Members}
\tablehead{
 \colhead{Group Redshift} &
 \colhead{Galaxy Name} &
 \colhead{R.A.} &
 \colhead{Dec} &
 \colhead{$z$} &
 \colhead{$i$\tablenotemark{a}} &
 \colhead{$R$\tablenotemark{b}}
}
\startdata
0.716 &  J152141.516+525422.87  &  15 21 41.516  &  52 54 22.870  &  0.7154  &   21.317 &  21.806 \\
 &  J152201.879+525154.94  &  15 22 01.879  &  52 51 54.943  &  0.7159  &   20.258 & \nodata \\
 &  J152125.755+525626.37  &  15 21 25.755  &  52 56 26.366  &  0.7166  &   20.670 & \nodata \\
 &  J152154.236+525212.74  &  15 21 54.236  &  52 52 12.737  &  0.7166  &  \nodata &  23.704 \\
 &  J152157.414+525140.39  &  15 21 57.414  &  52 51 40.395  &  0.7170  &  \nodata &  23.955 \smallskip \\
0.758 &  J152149.639+525357.38  &  15 21 49.639  &  52 53 57.385  &  0.7539  &   21.180 &  22.174 \\
 &  J152128.545+525631.31  &  15 21 28.545  &  52 56 31.306  &  0.7539  &   21.450 &  22.145 \\
 &  J152130.843+525642.03  &  15 21 30.843  &  52 56 42.026  &  0.7548  &   20.740 &  21.215 \\
 &  J152152.978+525456.06  &  15 21 52.978  &  52 54 56.062  &  0.7572  &   21.494 &  22.418 \\
 &  J152136.775+525504.75  &  15 21 36.775  &  52 55 04.751  &  0.7573  &   21.562 &  22.357 \\
 &  J152133.004+525614.24  &  15 21 33.004  &  52 56 14.241  &  0.7575  &   21.118 &  22.169 \\
 &  J152148.084+525229.23  &  15 21 48.084  &  52 52 29.227  &  0.7590  &  \nodata &  23.202 \\
 &  J152130.631+525643.55  &  15 21 30.631  &  52 56 43.555  &  0.7593  &   20.882 &  21.878 \\
 &  J152140.236+525420.07  &  15 21 40.236  &  52 54 20.069  &  0.7594  &  \nodata &  23.251 \\
 &  J152128.163+525657.95  &  15 21 28.163  &  52 56 57.952  &  0.7594  &   21.088 & \nodata \\
 &  J152133.198+525640.07  &  15 21 33.198  &  52 56 40.068  &  0.7599  &   20.984 &  22.100 \\
 &  J152144.852+525448.52  &  15 21 44.852  &  52 54 48.521  &  0.7604  &   17.995\tablenotemark{c} &  18.071\tablenotemark{c} \\
 &  J152139.952+525435.53  &  15 21 39.952  &  52 54 35.532  &  0.7630  &  \nodata &  23.862 \smallskip \\
0.818 &  J152139.850+525524.08  &  15 21 39.850  &  52 55 24.082  &  0.8132  &   22.137 &  23.352 \\
 &  J152205.728+525538.52  &  15 22 05.728  &  52 55 38.522  &  0.8168  &   21.180 &  21.663 \\
 &  J152153.761+525317.33  &  15 21 53.761  &  52 53 17.327  &  0.8172  &   20.504 &  22.846 \\
 &  J152148.911+525708.53  &  15 21 48.911  &  52 57 08.526  &  0.8175  &   21.901 &  23.341 \\
 &  J152128.683+525352.65  &  15 21 28.683  &  52 53 52.653  &  0.8178  &  \nodata &  23.690 \\
 &  J152125.164+525411.23  &  15 21 25.164  &  52 54 11.229  &  0.8178  &  \nodata &  23.096 \\
 &  J152130.942+525514.19  &  15 21 30.942  &  52 55 14.195  &  0.8182  &   21.547 &  21.961 \\
 &  J152146.267+525547.69  &  15 21 46.267  &  52 55 47.690  &  0.8185  &   20.915 &  21.823 \\
 &  J152154.462+525233.71  &  15 21 54.462  &  52 52 33.711  &  0.8197  &  \nodata &  23.733 \\
 &  J152139.973+525532.23  &  15 21 39.973  &  52 55 32.229  &  0.8226  &  \nodata &  23.011 \\
\enddata
\tablenotetext{a}{From SDSS imaging \citep{sdss}.}
\tablenotetext{b}{From J.~P.~Kneib, private communication.}
\tablenotetext{c}{Magnitudes include emission from the lensed quasar.}
\label{table_group_members}
\end{deluxetable*}

\begin{figure*}[ht]
\begin{center}
 \includegraphics[width=0.4\textwidth,clip=True]{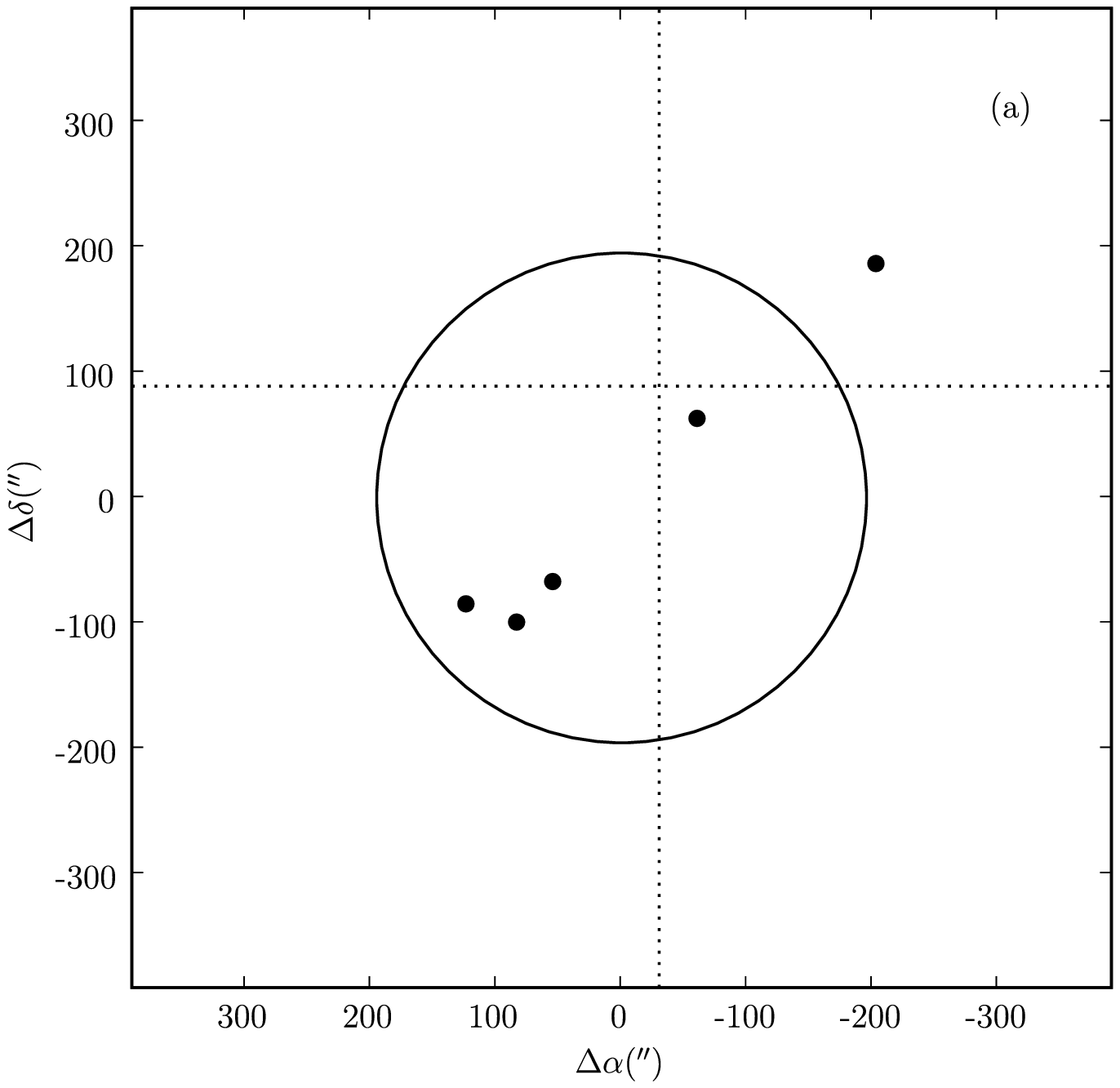}
 \includegraphics[width=0.4\textwidth,clip=True]{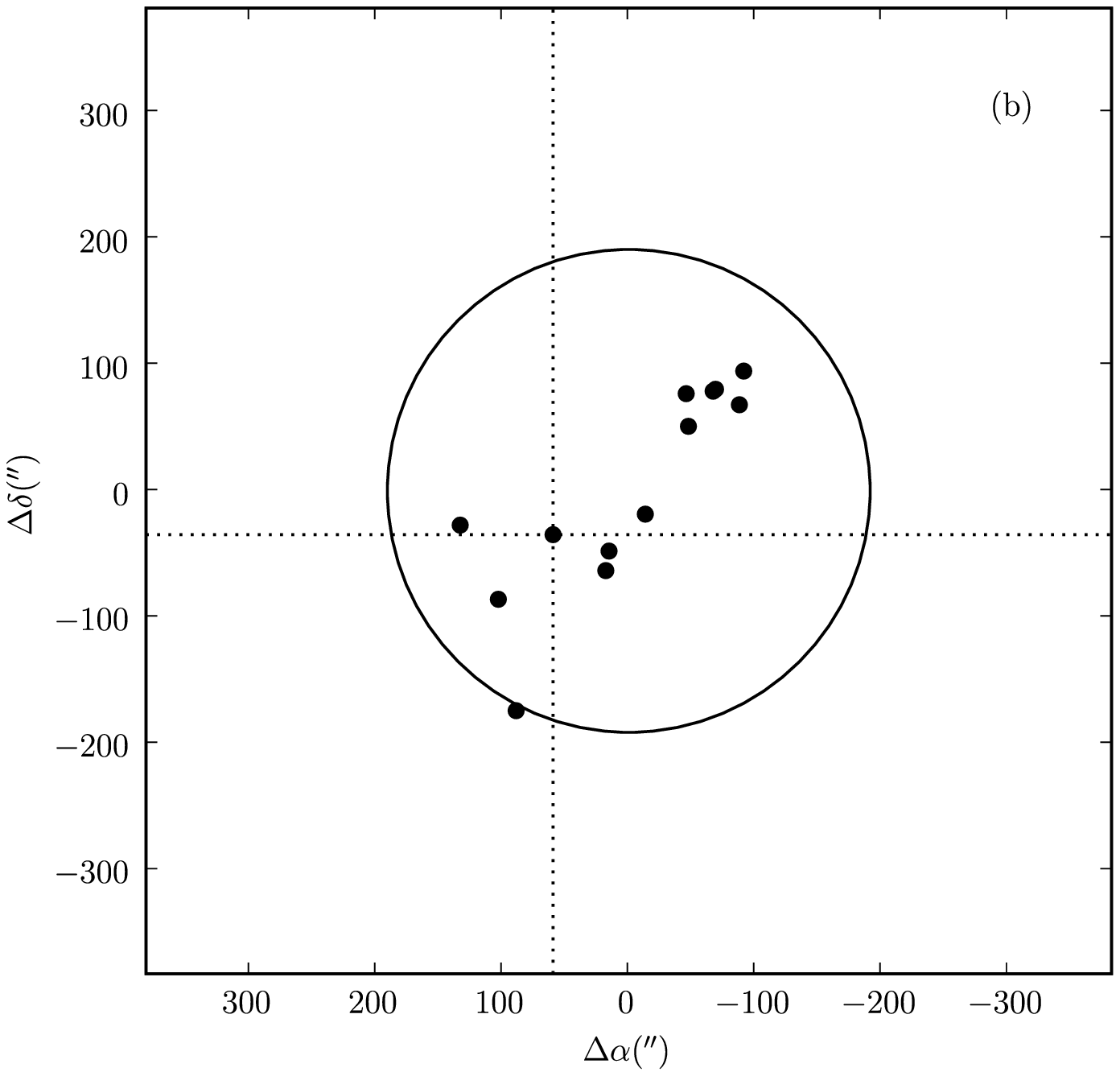}
 \includegraphics[width=0.4\textwidth,clip=True]{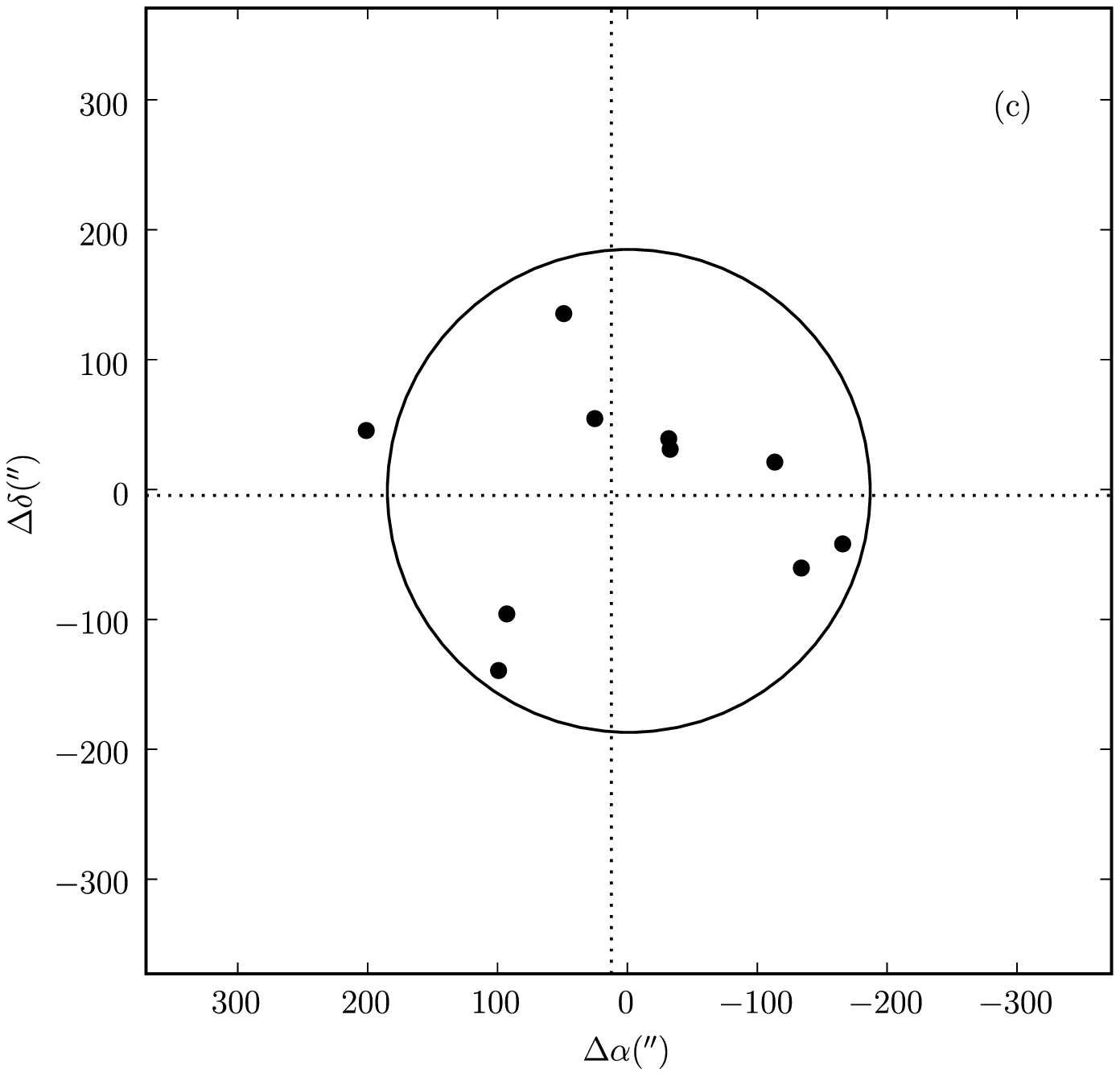}
  \caption{The spatial distribution of galaxies in the groups at (a) $z = 0.716$, (b) $z = 0.758$, and (c) $z = 0.818$. The location of the lens is marked with a dashed crosshair and the circle is centered on the group position and has a radius of 1~\hinv~Mpc at the redshift of the group.}
\label{figure_positions}
\end{center}
\end{figure*}

The spatial distribution of these systems is shown in Figure \ref{figure_positions}. Note that our data have a paucity of spectra in the immediate vicinity of the lens due to several factors. First, both of the slitmasks that we used in the observations contained a slit at least 10\arcsec~long placed over the lensing galaxy and oriented along the axis joining the two quasars, precluding the placement of slits across immediately neighboring galaxies. Additionally, the masks were approximately centered on the lens and therefore the LRIS slitmask bar obscures galaxies near the lens system. Finally, both masks placed a slit over a very bright star near the lens to help with point spread function modeling instead of placing the slit on neighbor galaxies. Therefore, the redshifts of galaxies very near to the lens were not sampled and this may systematically increase our estimate of the group offsets from the lens center (and thereby decrease the estimates of $\kappa$).

\section{DISCUSSION}
\subsection{The SBS1520 Lens Redshift}
The appearance of two sets of stellar absorption lines in the quasar spectra indicates that the lens must be modeled with at least two galaxy components \citep[e.g.,][]{burud}. However, it is not clear at which redshift the lensing galaxy is located or which of the neighboring galaxies could be assigned the redshift of the second absorption system. The presence of stronger absorption in the B component compared to the A component for the $z = 0.761$ system, the relative strengths of the stellar CaII features between the $z = 0.717$ system and the $z = 0.761$ system, and the relative sizes of the groups at $z = 0.716$ and $z = 0.758$ favor the interpretation that the lens is associated with the $z = 0.761$ absorption system, though the data are still inconclusive.

\subsection{Lensing Effects from the Environment}
We model the groups as SIS mass distributions and find the total group contribution to the convergence to be $\kappa_{grps} \sim 0.08$ (Table \ref{table_lens_groups}). The addition of more group members could increase or decrease the group velocity dispersion estimate, and we therefore assume that our current data provide a reasonable estimate of the true velocity dispersion of the groups. The biased sampling of group members (i.e., excluding potential members close to the lens) further complicates our ability to assess the group contributions to the lensing. The lack of spectral coverage around the lens would tend to inflate our estimate of the group offsets from the lens; the convergences reported in Table \ref{table_lens_groups} may therefore be underestimates because $\kappa_{SIS} \propto 1/r$. However, neither the lens nor any galaxies excluded from our spectroscopy are candidate brightest group galaxies that might be expected to lie near the center of the group potential; we therefore do not expect this bias to greatly affect our results.

In previous work we have shown \citep{auger} that the individual galaxies tend to provide a much larger contribution to the environmental convergence than the group halo if the lens is not the brightest group galaxy, as is the case with SBS1520. Therefore, all galaxies within a projected distance of several truncation radii from the lens need to be accounted for to adequately model the lens environment. Our \emph{HST} imaging shows 5 galaxies within a projected distance of 80~\hinv~kpc (for $z = 0.761$) of the lensing galaxy (Figure \ref{figure_f814w}), and the colors of these galaxies are consistent with being at approximately the same redshift as the lens galaxy. Four of these galaxies have not been treated in previous lens models, and assigning all four galaxies a redshift $z = 0.76$ and velocity dispersion $\sigma = 120~\mkms$ yields an additional convergence of $\kappa_{gals} \sim 0.11$ for SIS profiles. Previous lens models for SBS1520 have included cluster halos in the modeling \citep{burud}, but the contribution from these four individual galaxies was not accounted for in those models and can therefore be treated as a correction to previous models.

\subsection{Modeling the Mass Slope of SBS1520}
Early-type galaxies have been shown to have nearly isothermal mass profiles with only a small intrinsic scatter \citep{koopmans} and it is therefore reasonable to use SIS mass distributions to model lenses with a small number of constraints. However, part of the scatter in the \citet{koopmans} result may be due to systematic effects \citep{augerslacs} and our improved knowledge of the lens environment and the ellipticity of the lens galaxy makes it worthwhile to create updated lens models for the system that do not require the assumption of an isothermal mass profile. \citet{dobke} have used N-body simulations to suggest that galaxies in overdense environments may undergo stripping that creates steepened mass slopes when interacting with other galaxies. This presents a physical motivation for allowing for non-isothermal profiles in lens modeling and may explain why some lenses have systematically underestimated $H_0$ \citep[e.g.,][]{koopmans1600,treu1115,auger}. Considering that several independent estimates of $H_0$ have converged to approximately the same value of approximately 72\hunit, we have chosen to fix the value of $H_0$ at 72\hunit and allow the mass slope to vary in our model fitting.

We model the galaxy groups as SIS halos with the velocity dispersions and positions reported in Table \ref{table_lens_groups} and we model all galaxies within $\sim 80~$\hinv~kpc (except Galaxy M) as SIS halos with velocity dispersion $\sigma = 120~\mkms$. We model Galaxy M as an SIS halo but leave the velocity dispersion as a free parameter. The lens is modeled with {\em lensmodel} \citep{keeton} as a power-law mass distribution, $\rho \propto r^{-\alpha}$, and the slope and lens strength are allowed to vary in the optimization; the ellipticity, position angle, and location of the lens are fixed to the values obtained from our AO imaging. The time delay is $\Delta t = 130\pm3$ days \citep{burud} and, in principle, the flux ratio is best determined by using the time-delay corrected ratio from monitoring campaigns \citep[A/B = 1.89,][]{burud}. However, \citet{burud} do not perform their modeling with this value and do not directly comment on why the time-delay flux ratio was ignored; we were also unable to produce models of the lens that converged with such a low flux ratio. We use the quasar MgII emission line from our LRIS data to determine the flux ratio to be $2.9\pm0.4$, where the error is derived from the scatter between the seven independent LRIS spectra. We choose the quasar MgII line because it should be relatively insensitive to quasar variability and it is the reddest line available, thus minimizing problems with differential extinction \citep[e.g.,][]{wisotzki}.

SBS1520 can be reasonably well modeled with a steeper than isothermal power law profile ($\alpha = 2.29^{+0.08}_{-0.11}$) if $H_0$ is set fixed to 72\hunit (Table \ref{table_lens_parameters}); with three degrees of freedom (DOF), $\chi^2/{\rm DOF} = 1.1$. The main contribution to the chi-square comes from the flux ratio; the lens favors a flux ratio higher than the ratio derived from our data. We note that \citet{burud} used a flux ratio of 3.6 derived from emission lines in their spectra, though this is substantially higher than the emission line ratio from our spectroscopic data and the continuum flux ratio of 2.4 that we find in our NIRC2 K-band imaging.

\begin{deluxetable*}{ll}
\tabletypesize{\scriptsize}
\tablecolumns{2}
\tablewidth{0pc}
\tablecaption{Lens Model Parameters}
\tablehead{
 \colhead{Parameter} &
 \colhead{Value}
}
\startdata
Lens Position & -1\farcs1531, -0\farcs4150 \\
QSO A Position & 0\farcs0, 0\farcs0 \\
QSO B Position & -1\farcs4227, -0\farcs6569 \\
Flux Ratio (A/B) & 2.9 \\
$\Delta$t & 130 days \\
$H_0$ & 72\hunit \\
Power Law Index\tablenotemark{a} & 2.29 \\
Lens Strength\tablenotemark{a} & 0.53 \\
$\sigma_{\rm{Galaxy M}}$\tablenotemark{a} & 84\kms \\
$\chi^2$ & 3.4 \\
\enddata
\tablenotetext{a}{Parameter was varied in the model fitting.}
\label{table_lens_parameters}
\end{deluxetable*}

There is a well-known degeneracy between the mass slope and $H_0$ \citep[e.g.,][]{williams,wucknitz} that allows a substantial amount of play between these parameters \citep[compare our results with][for example]{burud}. The value of $H_0$ obtained for SBS1520 for an isothermal mass distribution when the environment is fully taken into account may be as low as $H_0 \approx 46$\hunit if we account for the unmodeled convergence due to the environment, $\kappa_{ext}$, via the simple relation
$$
H_{0, true} = (1 - \kappa_{ext}) H_{0, model}
$$
and use the isothermal models of \citet{burud}. The mass slope modeled in our analysis is steeper than expected for early-type galaxies \citep[e.g.,][]{koopmans}, but the presence of a nearby galaxy, Galaxy M, suggests that SBS1520 may be undergoing interaction-induced steepening \citep[e.g.,][]{dobke}. \citet{read} have used pixelated mass models to model SBS1520 and find a nearly isothermal profile, $\alpha = 1.95$. These models are able to account for `shape degeneracies' that may also effect interpretations of time delays \citep{saha}, though the analysis of \citet{saha} indicates that shape degeneracies are probably not important for SBS1520. Note that \citet{read} do not explicitly model the lens environment and they are therefore measuring the joint projected mass slope of the lens and environment. If the lens is not near the (projected) group center, th inferred mass slope would be shallower than the true mass slope.

\section{CONCLUSIONS}
We have obtained deep AO imaging and optical spectroscopy of the time-delay lens SBS1520. The AO imaging has allowed us to fix the lens galaxy properties with a high degree of precision when performing the lens modeling, and the data indicate that the lens has an elliptical morphology and perhaps a disk. The new spectroscopic data suggest that previous determinations of the lens redshift may be incorrect, and the data also allow us to quantify the lensing contribution of several groups in the immediate foreground and background of the lens. Lens models created with these new data can be well-fit with a steeper than isothermal mass slope ($\alpha = 2.29$) if $H_0$ is fixed at 72\hunit; isothermal models require $H_0 \sim 50$\hunit. \citet{dobke} found that galaxies in overdense environments might have steeper than isothermal mass slopes caused by interactions with other galaxies \citep[also see][]{augerslacs}. This suggests an interpretation that we are observing transient steepening of the mass profile due to galaxy-galaxy interactions and indicates that other lens systems that have obtained anomalously low values of $H_0$ may lie in overdense regions and near an interacting galaxy \citep[e.g., B1600+434;][]{koopmans1600,auger}. Alternatively, SBS1520 can be modeled in a manner consistent with an isothermal profile and $H_0 = 64$\hunit~if if the lens is modeled by a pixelated mass distribution and jointly modeled with other lens systems \citep{read}. These models indicate that twisting ellipticity, triaxial projection effects, or other shape degeneracies may be effecting the parametric analyses of SBS1520 \citep{saha}.

However, there are still several ambiguities in the data that need to be resolved before making definitive claims about the profile of SBS1520, particularly in the context of the interaction-induced steepening scenario. While we have argued that the lens redshift is likely to be $z = 0.761$ and not $z = 0.717$, the data are not conclusive. Furthermore, our modeling has assumed that all of the neighbor galaxies are at the group redshift; if this is the case, the $z = 0.76$ group centroid would be pulled closer to the lens and the group would therefore provide a larger contribution to the lens model. If this is not the case, the neighboring galaxies might have a smaller impact on the lens model. This is particularly important for Galaxy M, as this is the neighboring galaxy that most affects the lens model but also has colors least like the lens galaxy compared to the other field galaxies. It is also important to verify that at least one of the neighboring galaxies is at the same redshift as the lens because this is a requirement of the interaction-driven steepening hypothesis. Finally, obtaining a dynamical estimate of the lens mass would help to further constrain models and potentially distinguish between shape degeneracies and the mass slope degeneracy.


\acknowledgments 
We thank Lori Lubin, David Rusin, and John McKean for useful discussions and helpful comments. We also thank the referee for helpful suggestions. This work is based in part on observations made with the NASA/ESA Hubble Space Telescope, obtained from the the Data Archive at the Space Telescope Science Institute (STScI). STScI is operated by the Association of Universities for Research in Astronomy, Inc., under NASA contract NAS5-26555. These observations are associated with program \#AR-10300, supported by NASA through a grant from STScI. Some of the data presented herein were obtained at the W.M. Keck Observatory, which is operated as a scientific partnership among the California Institute of Technology, the University of California and the National Aeronautics and Space Administration. The Observatory was made possible by the generous financial support of the W.M. Keck Foundation. The authors wish to recognize and acknowledge the very significant cultural role and reverence that the summit of Mauna Kea has always had within the indigenous Hawaiian community.  We are most fortunate to have the opportunity to conduct observations from this mountain. This work has made use of the SDSS database. Funding for the SDSS and SDSS-II has been provided by the Alfred P. Sloan Foundation, the Participating Institutions, the National Science Foundation, the U.S. Department of Energy, the National Aeronautics and Space Administration, the Japanese Monbukagakusho, the Max Planck Society, and the Higher Education Funding Council for England. Part of this work was supported by the European Community's Sixth Framework Marie Curie Research Training Network Programme, Contract No. MRTN-CT-2004-505183 ``ANGLES".


\clearpage


\newpage

\begin{thebibliography}{}

\bibitem[Adelman-McCarthy et al.(2007)]{sdss} Adelman-McCarthy, J.~K., \& for  
the SDSS Collaboration 2007, ArXiv e-prints, 707, arXiv:0707.3413

\bibitem[Auger et al.(2007)]{auger} Auger, M.~W., Fassnacht, C.~D., Abrahamse, A.~L., Lubin, L.~M., \& Squires, G.~K.\ 2007, \aj, 134, 668

\bibitem[Auger (2007)]{augerslacs} Auger, M.~W.\ 2007, \mnras, in press (arXiv:0710.1651)

\bibitem[Bertin \& Arnouts(1996)]{bertin} Bertin, E., \& Arnouts, S.\ 1996, \aaps, 117, 393

\bibitem[Burud et al.(2002)]{burud} Burud, I., et al.\ 2002, \aap, 391, 481

\bibitem[Chavushyan et al.(1997)]{chavushyan} Chavushyan, V.~H., Vlasyuk, V.~V., Stepanian, J.~A., \& Erastova, L.~K.\ 1997, \aap, 318, L67

\bibitem[Churchill et al.(2005)]{churchill} Churchill, C.~W., Kacprzak, G.~G., \& Steidel, C.~C.\ 2005, IAU Colloq.~199: Probing Galaxies through Quasar Absorption Lines, 24

\bibitem[Courbin et al.(2000)]{courbin} Courbin, F., Magain, P., Kirkove, M., \& Sohy, S.\ 2000, \apj, 529, 1136

\bibitem[Crampton et al.(1998)]{crampton} Crampton, D., Schechter, P.~L., \& Beuzit, J.-L.\ 1998, \aj, 115, 1383

\bibitem[Dobke et al.(2007)]{dobke} Dobke, B.~M., King, L.~J., \& Fellhauer, M.\ 2007, \mnras, 377, 1503

\bibitem[Eigenbrod et al.(2007)]{eigenbrod} Eigenbrod, A., Courbin, F., \& Meylan, G.\ 2007, \aap, 465, 51

\bibitem[Faure et al.(2002)]{faure} Faure, C., Courbin, F., Kneib, J.~P., Alloin, D., Bolzonella, M., \& Burud, I.\ 2002, \aap, 386, 69

\bibitem[Freedman et al.(2001)]{freedman} Freedman, W.~L., et al.\ 2001, \apj, 553, 47

\bibitem[Fruchter \& Hook(2002)]{fruchter02} Fruchter, A.~S., \& Hook, R.~N.\ 2002, \pasp, 114, 144

\bibitem[Gaynullina et al.(2005)]{gaynullina} Gaynullina, E.~R., et al.\ 2005, \aap, 440, 53

\bibitem[Keeton(2001)]{keeton} Keeton, C.~R.\ 2001, ArXiv Astrophysics e-prints, arXiv:astro-ph/0102340

\bibitem[Kelson(2003)]{kelson} Kelson, D.~D.\ 2003, \pasp, 115, 688

\bibitem[Khamitov et al.(2006)]{khamitov} Khamitov, I.~M., Bikmaev, I.~F., Aslan, Z., Sakhibullin, N.~A., Vlasyuk, V.~V., Zheleznyak, A.~P., \& Zakharov, A.~F.\ 2006, Astronomy Letters, 32, 514

\bibitem[Koekemoer et al.(2002)]{koekemoer} Koekemoer, A.~M., Fruchter, A.~S., Hook, R.~N., \& Hack, W.\ 2002, The 2002 HST Calibration Workshop, p.337

\bibitem[Koopmans et al.(2000)]{koopmans1600} Koopmans, L.~V.~E., de Bruyn, A.~G., Xanthopoulos, E., \& Fassnacht, C.~D.\ 2000, \aap, 356, 391

\bibitem[Koopmans et al.(2003)]{koopmans03} Koopmans, L.~V.~E., Treu, T., Fassnacht, C.~D., Blandford, R.~D., \& Surpi, G.\ 2003, \apj, 599, 70

\bibitem[Koopmans et al.(2006)]{koopmans} Koopmans, L.~V.~E., Treu, T., Bolton, A.~S., Burles, S., \& Moustakas, L.~A.\ 2006, \apj, 649, 599

\bibitem[Oke(1990)]{hz44} Oke, J.~B.\ 1990, \aj, 99, 1621

\bibitem[Oke et al.(1995)]{oke} Oke, J.~B., et al.\ 1995, \pasp, 107, 375

\bibitem[Read et al.(2007)]{read} Read, J.~I., Saha, P., \& 
Macci{\`o}, A.~V.\ 2007, \apj, 667, 645 

\bibitem[Rusin et al.(2003)]{rusin} Rusin, D., Kochanek, C.~S., \& Keeton, C.~R.\ 2003, \apj, 595, 29

\bibitem[Saha \& Williams(2006)]{saha} Saha, P., \& 
Williams, L.~L.~R.\ 2006, \apj, 653, 936 

\bibitem[Sheinis et al.(2002)]{sheinis} Sheinis, A.~I., Bolte, M., Epps, H.~W., Kibrick, R.~I., Miller, J.~S., Radovan, M.~V., Bigelow, B.~C., \& Sutin, B.~M.\ 2002, \pasp, 114, 851

\bibitem[Spergel et al.(2007)]{spergel} Spergel, D.~N., et al.\ 2007, \apjs, 170, 377

\bibitem[Treu \& Koopmans(2002)]{treu1115} Treu, T., \& 
Koopmans, L.~V.~E.\ 2002, \mnras, 337, L6

\bibitem[Williams \& Saha(2000)]{williams} Williams, L.~L.~R., \& Saha, P.\ 2000, \aj, 119, 439

\bibitem[Wisotzki et al.(2003)]{wisotzki} Wisotzki, L., Becker, T., Christensen, L., Helms, A., Jahnke, K., Kelz, A., Roth, M.~M., \& Sanchez, S.~F.\ 2003, \aap, 408, 455

\bibitem[Wucknitz(2002)]{wucknitz} Wucknitz, O.\ 2002, \mnras, 332, 951

\bibitem[York et al.(2005)]{york} York, T., Jackson, N., Browne, I.~W.~A., Wucknitz, O., \& Skelton, J.~E.\ 2005, \mnras, 357, 124

\end{thebibliography}
\end{document}